\documentclass[sigconf]{acmart}

\usepackage{xcolor}
\usepackage{amsmath}
\usepackage{multirow}
\usepackage{graphicx}
\usepackage{subfigure}
\usepackage{colortbl}
\usepackage{makecell}
\usepackage{float}
\usepackage{enumitem}
\AtBeginDocument{%
  \providecommand\BibTeX{{%
    \normalfont B\kern-0.5em{\scshape i\kern-0.25em b}\kern-0.8em\TeX}}}

\begin{document}

\title{Breaking the Length Barrier: LLM-Enhanced CTR Prediction in Long Textual User Behaviors}

\author{Binzong Geng}
\authornote{Both authors contributed equally to this research.}
\affiliation{%
  \institution{Ant Group}
  \city{Hangzhou}
  \country{China}}
\email{gengbinzong.gbz@antgroup.com}

\author{Zhaoxin Huan}
\authornotemark[1]
\affiliation{%
  \institution{Ant Group}
  \city{Hangzhou}
  \country{China}}
\email{zhaoxin.hzx@antgroup.com}



\author{Xiaolu Zhang}
\affiliation{%
  \institution{Ant Group}
  \city{Hangzhou}
  \country{China}}
  \email{yueyin.zxl@antfin.com}

\author{Yong He}
\affiliation{%
  \institution{Ant Group}
  \city{Hangzhou}
  \country{China}}
  \email{heyong.h@antgroup.com}
  

\author{Liang Zhang}
\affiliation{%
  \institution{Ant Group}
  \city{Hangzhou}
  \country{China}}
  \email{zhuyue.zl@antfin.com}

\author{Fajie Yuan}
\affiliation{%
  \institution{Westlake University}
  \city{Hangzhou}
  \country{China}}
  \email{yuanfajie@westlake.edu.cn}

\author{Jun Zhou}
\authornote{Corresponding author.}
\affiliation{%
  \institution{Ant Group}
  \city{Hangzhou}
  \country{China}}
\email{jun.zhoujun@antfin.com}

\author{Linjian Mo}
\authornotemark[2]
\affiliation{%
  \institution{Ant Group}
  \city{Hangzhou}
  \country{China}}
  \email{linyi01@antgroup.com}
\begin{abstract}
With the rise of large language models (LLMs), recent works have leveraged LLMs to improve the performance of click-through rate (CTR) prediction. However, we argue that a critical obstacle remains in deploying LLMs for practical use: the efficiency of LLMs when processing long textual user behaviors.
As user sequences grow longer, the current efficiency of LLMs is inadequate for training on billions of users and items.
To break through the efficiency barrier of LLMs, we propose Behavior Aggregated Hierarchical Encoding (BAHE) to enhance the efficiency of LLM-based CTR modeling.
Specifically, BAHE proposes a novel hierarchical architecture that decouples the encoding of user behaviors from inter-behavior interactions.
Firstly, to prevent computational redundancy from repeated encoding of identical user behaviors, BAHE employs the LLM's pre-trained shallow layers to extract embeddings of the most granular, atomic user behaviors from extensive user sequences and stores them in the offline database. 
Subsequently, the deeper, trainable layers of the LLM facilitate intricate inter-behavior interactions, thereby generating comprehensive user embeddings. This separation allows the learning of high-level user representations to be independent of low-level behavior encoding, significantly reducing computational complexity. Finally, these refined user embeddings, in conjunction with correspondingly processed item embeddings, are incorporated into the CTR model to compute the CTR scores.
Extensive experimental results show that BAHE reduces training time and memory by five times for CTR models using LLMs, especially with longer user sequences. BAHE has been deployed in a real-world system, allowing for daily updates of 50 million CTR data on 8 A100 GPUs, making LLMs practical for industrial CTR prediction.


\end{abstract}






\maketitle


\section{Introduction}
Click-through rate (CTR) prediction plays a significant role in various domains, including online advertising, search engines, and recommendation systems. Recently, large language models (LLMs) have achieved remarkable results across various domains~\cite{openai,fewshotllm,instructgpt,reasoning,survey,li2023exploring}. Numerous studies have investigated the application of LLMs in CTR prediction, leveraging LLMs' powerful semantic understanding and world knowledge to enhance CTR modeling.
For example, M6-Rec~\cite{m6} reconstructs the interactions between users and items in textual prompts and utilizes M6 to conduct several recommendation tasks including CTR prediction.
Both CTRL~\cite{ctrl} and FLIP~\cite{flip} integrate semantic information from LLMs into traditional ID-based models~\cite{yuan2023go} using contrastive learning and masked language modeling.
KAR~\cite{kar} develops a three-stage framework based on the reasoning and factual knowledge of LLMs, transferring knowledge from LLMs into the CTR model through embeddings.
These works demonstrate the tremendous potential and broad prospects for LLMs in enhancing CTR prediction.

However, we argue that a challenge for LLMs in practical CTR prediction remains unresolved: the efficiency bottleneck of LLMs when dealing with long user behavior sequences. This issue significantly hampers the real-world application of LLMs. 
In traditional CTR modeling, it is widely known that integrating diverse, long-term user sequences into the model can enhance its performance~\cite{din,ali_lifelong,ali_longseq,meituan}.
However,incorporating longer texts into LLM-based CTR models may improve performance but also introduces a bottleneck, as extended texts significantly slow down training and inference, making such models unsuitable for large-scale deployment.
This reason has led the LLM-based CTR methods mentioned before to compromise by using smaller language models and shorter user sequences~\cite{ctrl,roberta,flip,kar}.
Moreover, some LLM-based recommendation methods are not directly tailored to CTR tasks, such as sequential recommendation~\cite{taste,scaling,universal,p5,recformer,fu2023exploring,li2023exploring}. These methods are more focused on item matching or retrieval, where the user sequences only consist of items that have been interacted with. As a result, the issue of long user sequences is relatively less severe in these works.



To address these challenges, in this paper, we propose \underline{B}ehavior \underline{A}ggregated \underline{H}ierarchical \underline{E}ncoding (BAHE) to tackle the performance bottleneck of LLMs with long user sequences, facilitating the application of LLMs in real-world CTR prediction.
Specifically, BAHE proposes a novel hierarchical architecture that decouples the encoding of user behaviors from inter-behavior interactions.
Firstly, to understand the semantic meaning of each user behavior and to prevent computational redundancy from repeated encoding of identical user behavior, BAHE employs the LLM's pre-trained low layers to extract embeddings of the most granular, atomic user behaviors from extensive user sequences and stores them in the offline database.
In this way, BAHE converts the encoding from token-level to behavior-level, substantially reducing sequence length and enhancing the reusability of behavior.
Subsequently, BAHE retrieves all of a user's behaviors from the offline restored atomic behavior database and utilizes the deeper, trainable layers of the LLM to facilitate intricate inter-behavior interactions, thereby learning user preferences and generating comprehensive user embeddings.
This separation allows the learning of high-level user representations to be independent of low-level behavior encoding, significantly reducing computational complexity. Finally, these refined user embeddings, in conjunction with correspondingly processed item embeddings, are incorporated into the CTR model to compute the CTR scores.Significantly, atomic behaviors rarely undergo changes, which permits infrequent updates within the lower layers of the LLM. These updates can proceed asynchronously relative to those in the higher layers, thereby boosting computational efficiency. Furthermore, the input sequence for the LLM's higher layers is significantly compacted, reducing from the original token count to a smaller number of behaviors. This concise representation not only shortens the sequence length the LLM needs to process but also enhances the model's overall efficiency.

BAHE has been successfully implemented in real-world industrial CTR prediction, where the training time for LLM-based CTR with over 50 million data has been reduced from the initial 5 days to just 1 day, enabling daily model updates and scheduling. Overall, our main contributions are as follows:

\begin{itemize}[leftmargin=*]
    \item We investigate a significant and unresolved problem: the efficiency bottleneck in LLM-based CTR modeling with long user sequences. We find that the main reasons for the performance bottleneck due to the repetitive encoding of user behaviors and the strong coupling between behavior representation extraction and behavior interaction modeling.
    
    \item We propose the BAHE method, which is a novel hierarchical structure to LLM by decoupling the representation extraction of atomic behaviors from the learning of behavior interactions. BAHE solves redundant representations of the same behavior across different users and significantly reduces the length of input sequences.
    \item Extensive offline experiments validate that BAHE significantly enhances the efficiency of LLM-based CTR models. Moreover, online experiments fully demonstrate BAHE's ability to reduce computational resources in real-world industrial CTR prediction.
\end{itemize}



\section{Proposed Method}

\subsection{Problem Definition}\label{sec:problem_define}
LLM-based CTR prediction aims to estimate the probability of a user clicking on an item based on textual features. For each user $u \in U$, the user's textual behavior sequence is $s_u$, with the total token length of $s_u$ being $l_u$. $s_u$ is composed of $N$ behavior sequences from different domains (such as click, favorite, add-to-cart actions) denoted as $s_u = [s_{un} | 0 < n < N]$, with each sequence $s_{un}$ containing an average of $M$ user behaviors $s_{un} = [a_{unm} | 0 < m < M]$, and each behavior $a_{unm}$ having an average of $K$ tokens, which means $l_u = N \times M \times K$. For an item $i\in I$, the textual features of $i$ is $t_i$, comprised of the item's title, with the overall length $l_i$, which is significantly shorter than $l_u$. We define $H$ to be the set of distinct atomic behaviors across all users:

\begin{equation}
    H = \{a_{umn} | u \in U, 0 < m < M,  0 < n< N\}
\label{eq:H}
\end{equation}
Each user's behavior sequence $s_u$ consists of these atomic behaviors and includes the titles of items they've interacted with, hence $t_i \in H$. The objective of LLM-based CTR modeling is to minimize $L$:

\begin{equation}
    L = \frac{1}{|D|} \sum_{ u \in U, i \in I} l(Y, F(s_u,t_i))
\label{eq:loss}
\end{equation}
where $l$ represents the loss function, $|D|$ denotes the total number of training samples, and $Y$ indicates the click label.

\subsection{Behavior Aggregated Hierarchical Encoding}\label{sec:method}
Given users $i$ and $j$, with atomic behavior sequences  $s_i=[a_1, a_2, a_3]$ and $s_j=[a_3, a_1, a_2]$ respectively, Traditional LLM-based CTR modeling struggle with efficiency due to:

\begin{itemize}[leftmargin=*]
    \item \textbf{Redundant behavior encoding}: The same behaviors are redundantly encoded across different users' sequences. For example, $s_i$ and $s_j$ both contain $a_1$, $a_2$, and $a_3$, causing unnecessary repetition in encoding and a waste of computational resources.

    \item \textbf{Tight coupling}: Behaviors like $a_1$, $a_2$, and $a_3$ have fixed meanings, while their sequence varies per user. Existing approaches couple representation extraction with sequence understanding, causing regular, costly updates when behaviors change.
\end{itemize}

\begin{figure}[t]
  \centering
  \includegraphics[width=0.95\linewidth]{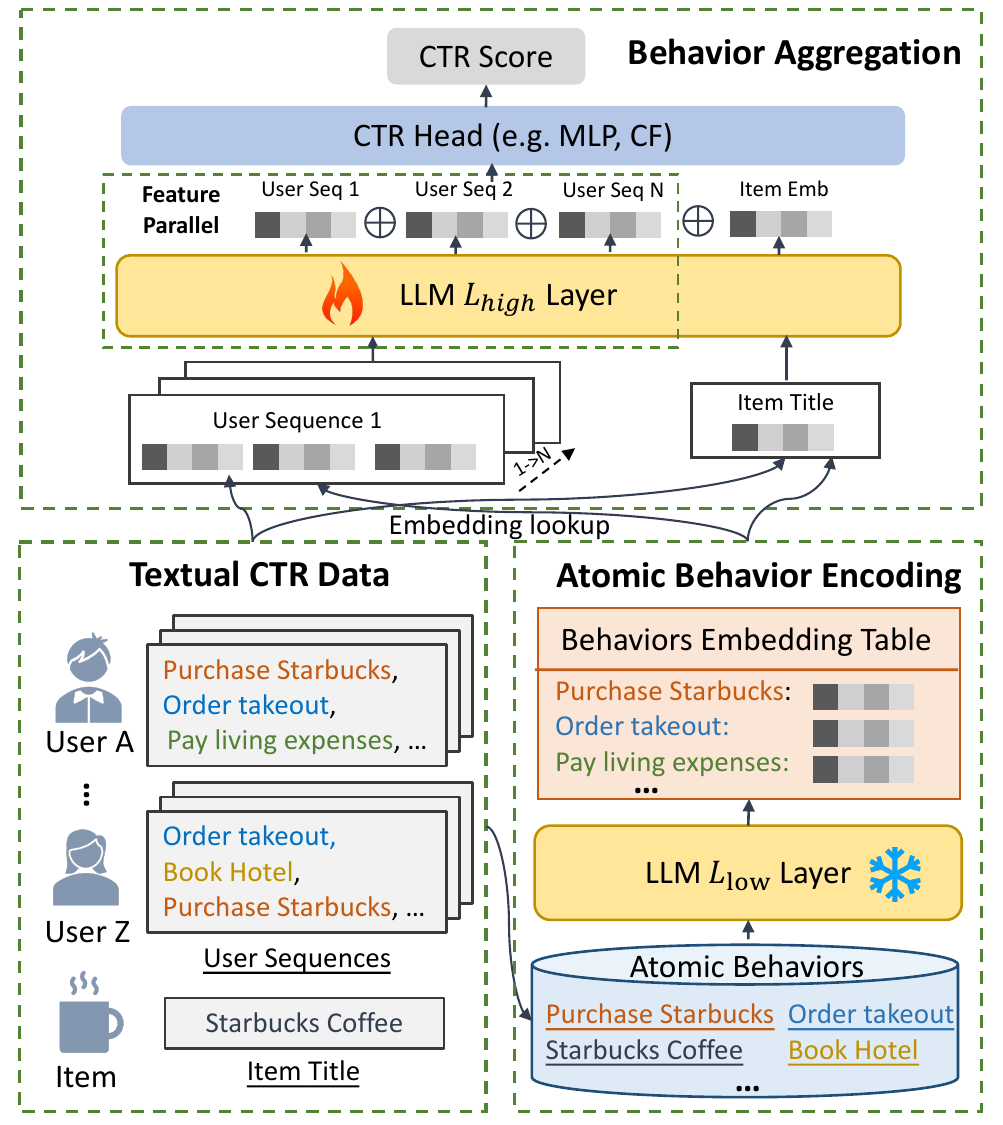}
  \caption{Architecture of the proposed BAHE method.}
  \label{fig:overview}
\end{figure}


To address these issues, we propose the Behavior Aggregated Hierarchical Encoding (BAHE) approach, and Figure~\ref{fig:overview} shows the architecture of BAHE. BAHE upgrades the LLM’s encoding from token-level to behavior-level, dramatically reducing the encoding length and increasing the cross-user reusability of behaviors. Additionally, BAHE stratifies the LLM and decouples representation extraction from behavior interaction, boosting computational efficiency and adaptability.

\subsubsection{Atomic Behavior Encoding (ABE)}
BAHE first encodes all behaviors from the set $H$ using the LLM's pre-trained low layers, denoted as $LLM_{L_{low}}$, then stores them offline to serve as a behavior embedding table $E$. Subsequently, the LLM's high layers will utilize the $E$ as a replacement for the original token embedding table to learn interactions among user behaviors. The encoding of atomic behaviors is illustrated as follows:


\begin{equation}
    E_{a_i} = F_{p}(LLM_{L_{low}}(a_i))  \quad E = \{a_i: E_{a_i} | a_i \in H\}
\label{eq:fpooling1}
\end{equation}
where $a_i$ is an atomic behavior defined in $H$, composed of $K$ text tokens. $LLM_{L_{low}}(a_i) \in R ^ {K \times d}$ corresponds to the $L_{low}$ hidden states, 
where $d$ is the dimension. $F_{p}: R^{K\times d} \Rightarrow R^{d}$ is the pooling function. The behavior embedding table is $E \in \mathbb{R}^{|H|\times d}$, where $|H|$ is the total number of atomic behaviors. In this way, BAHE transforms encoding from the token level to the behavior level, substantially reducing the encoding length from the number of tokens to the number of atomic behaviors. 


\subsubsection{Behavior Aggregation (BA)}
After obtaining the atomic behavior embedding table $E$, BAHE retrieves the corresponding $E_{a_i}$ for each $a_i$ in user $u$'s n-th sequence $s_{un}$, and then concatenates them as representation for $s_{un}$:

\begin{equation}
    E(s_{un}) = [E(a_{un1}) \oplus E(a_{un2}) \oplus \ldots \oplus E(a_{unM})]
\label{eq:E}
\end{equation}
BAHE next employs the LLM's higher layers $LLM_{L_{high}}$ to model the interaction between behaviors, and obtains the overall representation for user sequence $s_{un}$:
\begin{equation}
    Q_{un} = F_{d}(F_{p}(LLM_{L_{high}}(E(s_{un}))))
\label{eq:fpooling2}
\end{equation}
$Q_{un} \in \mathbb{R}^{d}$ is the representation of the n-th sequence. $F_{d}: R^{d} \Rightarrow R^{\hat{d}}$ is a dimensionality reduction function that aims to transform high-dimensional LLM hidden states into lower-dimensional, facilitating its use in subsequent models.
\subsubsection{Feature Parallel (FP)}
To avoid the exponential growth in LLM's attention computations as the number of user sequences $N$ increases, BAHE utilizes a parallel and independent way to process each user sequence through the $LLM_{L_{high}}$ and concatenate them to get the final user representation:
\begin{equation}
    Q_u = [Q_{u1} \oplus Q_{u2} \oplus \ldots \oplus Q_{uN}]
\end{equation}
Following the same method described above, BAHE can also obtain the representation of the item, denoted as $Q_i$.

\subsubsection{CTR Modeling}\label{sec:ctr}
After obtaining the user representation $Q_u$ and item representation $Q_i$, BAHE feeds their concatenation into the CTR model $F_{\theta}$ to compute the final CTR score $y$:

\begin{equation}
    y = F_{\theta}(Q_u \oplus Q_i) 
\end{equation}
BAHE is model-agnostic, which can apply to any embedding-based CTR model. For illustrative purposes, BAHE chose the simple but efficient DNN (Deep Neural Network) as default. BAHE optimizes the loss function defined in Equation~\ref{eq:loss} to learn the final LLM-based CTR model. After training, BAHE uses $Q_u$ and $Q_i$ to improve downstream model performance by providing additional semantic knowledge~\cite{kar}.



\subsection{Complexity Analysis}
We analyze the time complexity of BAHE against traditional LLM-based CTR models. With $N$ textual behavior features per user, $M$ behaviors per sequence, $H$ atomic actions, and $K$ tokens per action. The original complexity is $O(L(NMK)^2)$, where $L$ is the number of layers in the LLM. BAHE splits this into two stages: atomic behavior encoding $O(L_{low}(HK^2))$, and behavior aggregation with feature parallel $O(L_{high}(NM^2))$, where $L = L_{low} + L_{high}$. The efficiency improvement brought by BAHE is 
$O(L_{low}((N^2M^2-H)K^2)) + O(L_{high}((N^2K^2-N)M^2))$. 

BAHE simplifies the LLM's lower-level encoding, lowering complexity by $(N^2M^2-H)K^2$ as it encodes just $H$ atomic behaviors instead of redundantly processing $N^2M^2$ identical user behaviors. Furthermore, for LLM's higher-level semantic understanding, BAHE processes $N$ sequence features in parallel and employs behavior aggregation, cutting computations by $(N^2K^2-N)M^2$.

\section{Experiment}

\subsection{Experimental Setup}
\subsubsection{Dataset}
We evaluated BAHE using a real-world industrial dataset with around 50 million CTR records collected over a week.
The data, split into training, validation, and test sets by log time, includes 6 text features like user bills, searches, and mini-program visits, along with item titles. Each user sequence has 50 user behaviors, averaging 5 tokens each, summing up to 10 million atomic behaviors. The label indicates whether the user clicked on the item.


\subsubsection{Baseline Methods}
The baseline methods select mainstream LLM-based CTR modeling (denoted as LLM-CTR)~\cite{ctrl,kar,flip,unictr}, which concatenate multiple user text sequences into single long sequences, and then perform CTR modeling based on the LLM. 
Since BAHE is model-agnostic, it can be applied to any LLM-based models.Furthermore, to test the enhancement that LLM provides to the downstream CTR model, we also present the performance of the downstream CTR model, which is denoted as DNN.

\subsubsection{Evaluation Metrics}
We assess performance using the Area Under the Curve (AUC)~\cite{auc}, standard for CTR. Efficiency is evaluated by training GPU hours (GPU-h)~\cite{gpuh} and memory usage. Furthermore, as mentioned in Section~\ref{sec:ctr}, we also show AUC$_{d}$ to demonstrate LLM's benefits for downstream CTR models. 


\subsubsection{Experimental Details}

Our backbone LLM leverages the open-source Qwen-1\_8B\footnote{https://huggingface.co/Qwen/Qwen-1\_8B}~\cite{qwen}. As detailed in Section~\ref{sec:method}, we employ an MLP with [2048, 512, 128] units as $F_{d}$ to condense LLM embeddings, and another MLP as $F_{\theta}$ with [256, 32, 1] units for CTR predictions, using mean pooling as default $F_{p}$. We fine-tune using Lora Tuning~\cite{lora,fu2023exploring} at rank 64, batch size 16, one training epoch, and a 5e-5 learning rate with cosine decay. All tests run on 8 A100 GPUs.


\begin{table}[t]
\small
\setlength\tabcolsep{2.1pt}
  \centering
  \caption{Performance of BAHE at different text lengths. "AUC" represents the modeling performance of LLM, while "AUC$_{d}$" indicates the performance when transferring LLM representations to downstream models.}
    \begin{tabular}{c|ccc|cccc}
    \toprule
    \multirow{2}[4]{*}{Model} & \multicolumn{3}{c|}{\textbf{Text Length=1024}} & \multicolumn{4}{c}{\textbf{Text Length=2048}} \\
\cmidrule{2-8}          & AUC   & GPU-h & Mem(G) & AUC   & GPU-h & Mem(G) & AUC$_{d}$ \\
    \midrule
    \midrule
    DNN & - & -  & -  & - & -  & -  & 0.7219 \\
    LLM-CTR & 0.7161 & 448  & 43.8  & 0.7276 & 928  & 75.4  & 0.7323 \\
    + FP  & 0.7143 & 420  & 36.8  & 0.7326 & 864  & 67.8  & 0.7369 \\
    + ABE  & 0.7150 & 256   & 23.0    & 0.7332 & 416  & 38.0    & 0.7372 \\
    + BA(\textbf{BAHE}) & 0.7132 & 116   & 9.8   & 0.7309 & 164   & 12.6  & 0.7352 \\
    \bottomrule
    \end{tabular}%
  \label{tab:model performance}%
\end{table}%

\subsection{Performance}
\subsubsection{Offline Performance}
Table~\ref{tab:model performance} compares BAHE with the baseline. The findings are: Firstly, BAHE cuts training time by 5x, from 928 to 164 GPU hours, and slashes GPU memory to a sixth of the baseline. This marks a significant improvement in the efficiency of LLMs when processing long sequences. Secondly, BAHE's reuse of behavior representations improves key behavior capture, substantially raising AUC for both LLMs and downstream models. BAHE thus strikes an effective balance between efficiency and performance. Lastly, the individual components of BAHE-Feature Parallel (FP), Atomic Behavior Encoding (ABE), and Behavior Aggregation (BA)-contributes to these improvements, highlighting their importance. In summary, BAHE enhances both efficiency and effectiveness for LLMs in long sequence applications, paving the way for their practical use.

\subsubsection{Online Deployment}
To demonstrate BAHE's practical benefits, we deploy BAHE on a large-scale e-commerce platform's advertisement CTR prediction and conduct a two-week A/B test. BAHE allowed daily LLM training on 50 million CTR records, outperforming the baseline model's weekly training capacity. Consequently, BAHE unlocks a greater potential for LLM-based CTR, resulting in a 9.65\% increase in online CTR and 2.41\% rise in advertising CPM.

\begin{figure}[]
\centering
\subfigure[]{
\begin{minipage}[t]{0.22\textwidth}
\centering
\includegraphics[width=\textwidth]{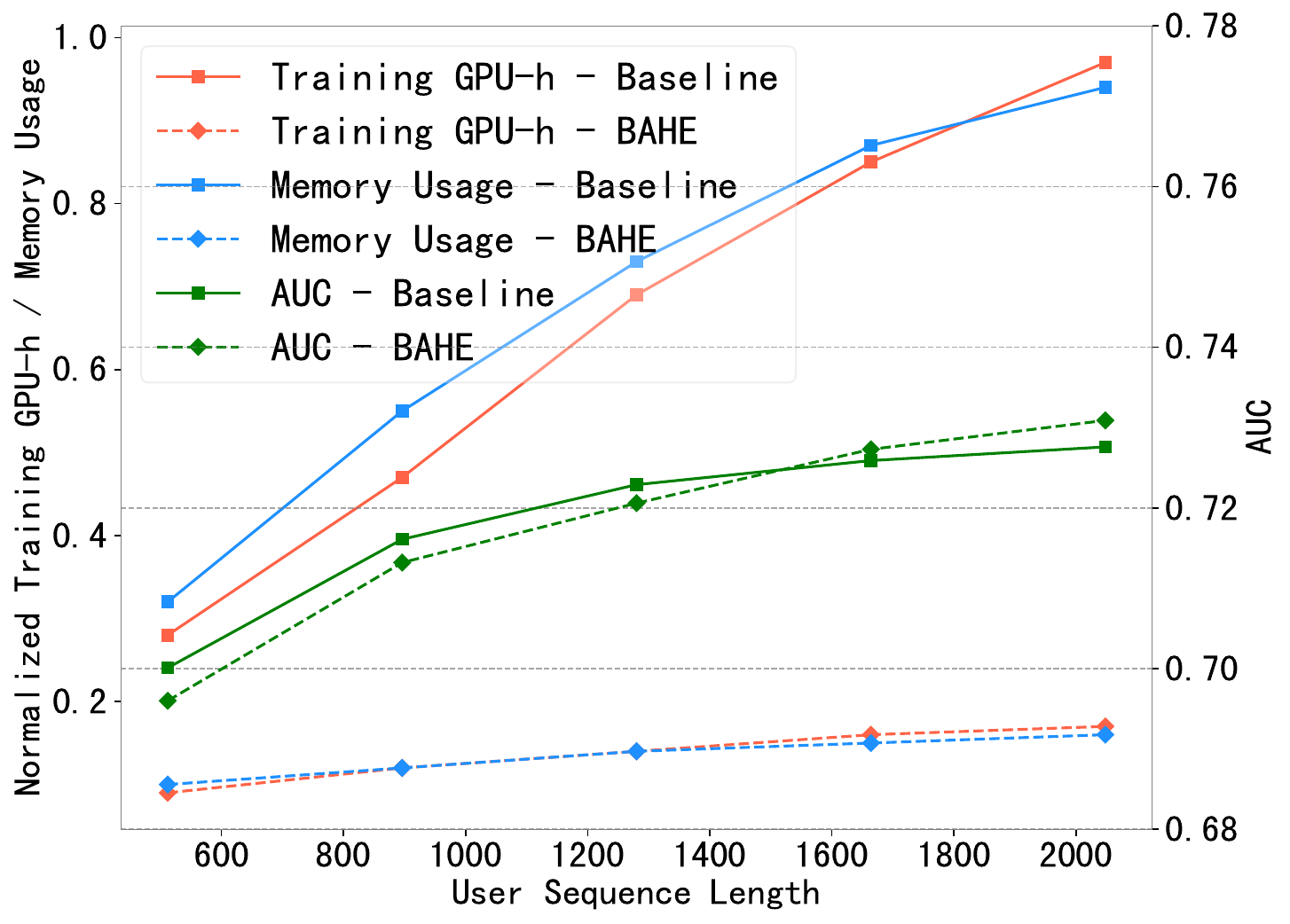}
\label{fig:length}
\end{minipage}
}
\subfigure[]{
\begin{minipage}[t]{0.22\textwidth}
\centering
\includegraphics[width=\textwidth]{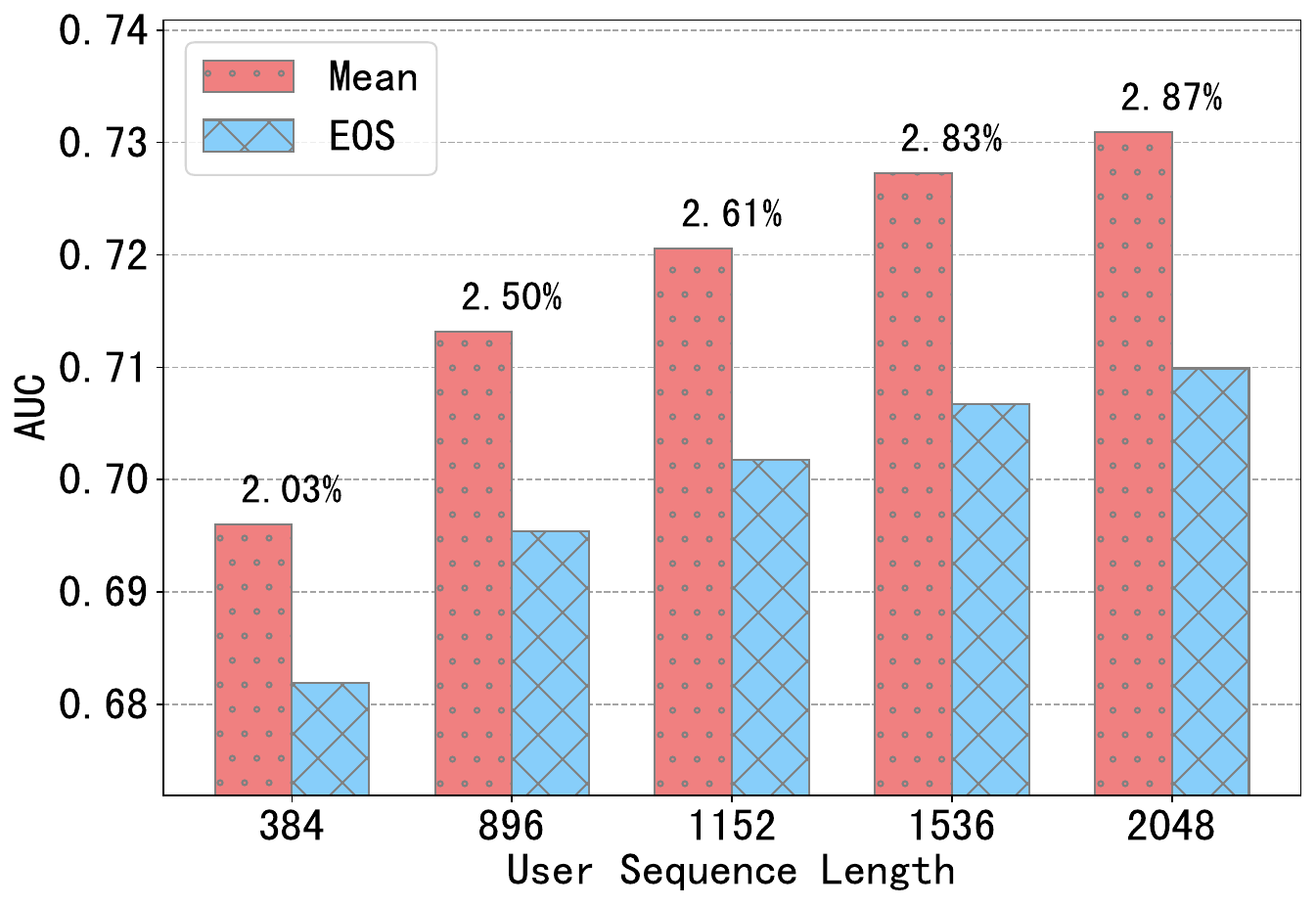}
\label{fig:pooling}
\end{minipage}
}
\caption{(a) presents a comparison of the normalized training time and memory usage (left y-axis) and the AUC (right y-axis) of BAHE with the baseline for different user sequence lengths (x-axis). Figure (b) shows the AUC comparison of BAHE utilizing different behavior pooling methods.
}
\end{figure}

\subsection{Empirical Analysis}
\subsubsection{Comparison at Different Sequence Lengths}
To give a more detailed analysis of the performance enhancements attributable to BAHE, Figure~\ref{fig:length} presents a comparison between BAHE and the baseline at different sequence lengths.
Firstly, AUC improves with longer user sequences, showing that extended texts enhance LLM-based CTR. Secondly, BAHE's performance boosts are greater for longer texts, indicating its effectiveness in handling increased lengths.



\subsubsection{Comparison of Different Pooling Methods}
During the extraction of atomic behavioral representations and subsequent behavior aggregation, the choice of pooling method (denoted as $F_{p}$ in Equations~\ref{eq:fpooling1} and~\ref{eq:fpooling2}) plays a crucial role. We evaluate two prevalent techniques in Figure~\ref{fig:pooling}: mean pooling and the last hidden state of the LLM (denoted as EOS).Our findings show mean pooling is superior to EOS, indicating global representations are more effective than the last hidden embedding for generative LLMs.



\section{Conclusion}
To tackle the challenge of efficiency in LLM-based CTR models when dealing with users' extensive text sequences, this paper proposes a novel BAHE method. BAHE enhances the reusability of behavioral representations across users by encoding atomic behaviors. It employs LLM's hierarchical encoding technique to separate the learning of behavior representations from the inter-behavior modeling, thereby boosting computational efficiency. Extensive online and offline experimental results demonstrate that BAHE not only achieves a more than 5 times increase in efficiency but also enhances the CTR performance, offering fresh insights for the practical deployment of LLM-based CTR models.

\bibliographystyle{ACM-Reference-Format}
\bibliography{bahe}

\end{document}